\documentclass[reprint,
superscriptaddress,
showpacs,preprintnumbers,
notitlepage,amsmath,amssymb,
prl,
]{revtex4-1}

\usepackage{graphicx}
\usepackage{dcolumn}
\usepackage{bm}
\usepackage{diagbox}

\usepackage{graphicx}
\usepackage{enumerate}
\usepackage{subfigure}
\usepackage{epstopdf}
\usepackage[colorlinks,
linkcolor=red,
anchorcolor=black,
citecolor=blue]{hyperref}
\usepackage{amssymb}
\usepackage{framed}
\usepackage{tabularx}
\usepackage{booktabs}
\usepackage{float}
\usepackage[table]{xcolor}
\usepackage{array}
\usepackage{tikz}
\hypersetup{colorlinks=true}

\begin{document}
\title{Construction and classification of point group symmetry-protected topological phases in 2D interacting fermionic systems}
\author{Jian-Hao Zhang}
\affiliation{Institute for Advanced Study, Tsinghua University, Beijing 100084, China}
\author{Qing-Rui Wang}
\affiliation{Department of Physics, The Chinese University of Hong Kong, Shatin, New Territories, Hong Kong, China}
\author{Shuo Yang}
\affiliation{State Key Laboratory of Low Dimensional Quantum Physics and Department of Physics, Tsinghua University, Beijing 100084, China}
\author{Yang Qi}
\email{qiyang@fudan.edu.cn}
\affiliation{Center for Field Theory and Particle Physics, Department of Physics, Fudan University, Shanghai 200433, China}
\affiliation{State Key Laboratory of Surface Physics, Fudan University, Shanghai 200433, China}
\affiliation{Collaborative Innovation Center of Advanced Microstructures, Nanjing 210093, China}
\author{Zheng-Cheng Gu}
\email{zcgu@phy.cuhk.edu.hk}
\affiliation{Department of Physics, The Chinese University of Hong Kong, Shatin, New Territories, Hong Kong, China}

\begin{abstract}
The construction and classification of symmetry-protected topological (SPT) phases in interacting bosonic and fermionic systems have been intensively studied in the past few years. Very recently, a complete classification and construction of space group SPT phases were also proposed for interacting bosonic systems. In this paper, we attempt to generalize this classification and construction scheme systematically into interacting fermion systems. In particular, we construct and classify point group SPT phases for 2D interacting fermion systems via lower-dimensional block-state decorations. We discover several intriguing fermionic SPT states that can only be realized in interacting fermion systems (i.e., not in free-fermion or bosonic SPT systems). Moreover, we also verify the recently conjectured crystalline equivalence principle for 2D interacting fermion systems. Finally, the potential experimental realization of these new classes of point group SPT phases in 2D correlated superconductors is addressed.
\end{abstract}

\maketitle
\newcommand{\lra}{\longrightarrow}
\newcommand{\xra}{\xrightarrow}
\newcommand{\ra}{\rightarrow}
\newcommand{\bs}{\boldsymbol}
\newcommand{\ul}{\underline}
\newcommand{\1}{\text{\uppercase\expandafter{\romannumeral1}}}
\newcommand{\2}{\text{\uppercase\expandafter{\romannumeral2}}}
\newcommand{\3}{\text{\uppercase\expandafter{\romannumeral3}}}
\newcommand{\4}{\text{\uppercase\expandafter{\romannumeral4}}}
\newcommand{\5}{\text{\uppercase\expandafter{\romannumeral5}}}
\newcommand{\6}{\text{\uppercase\expandafter{\romannumeral6}}}

\textit{Introduction} -- In recent years, the concept of quantum entanglement patterns has played an essential role in constructing and classifying topological phases of quantum matter. At a very basic level, the ground state of a gapped quantum system can be classified as a long-range or short-range entangled state. In the presence of global symmetry, even short-range entangled states can be classified into numerous different phases, including the symmetry-protected topological (SPT) phases~\cite{ZCGu2009,1D,XieChenScience,cohomology}, in addition to the conventional symmetry-breaking phases. The simplest example of an SPT phase is a topological insulator, which is protected by time-reversal and charge-conservation symmetry~\cite{KaneRMP,ZhangRMP}. Having a complete construction and classification of SPT phases is a crucial step towards understanding these peculiar quantum phases of matter. A general scheme of classifying bosonic SPT (BSPT) phases has been well established using group cohomology theory~\cite{XieChenScience,cohomology} and invertible topological quantum field theory (TQFT)~\cite{E8,invertible1,invertible2,invertible3}. An alternative strategy of classification was constructed in Ref.~\onlinecite{LevinGu} by ``gauging" the global symmetry and investigating the braiding statistics of gauge fluxes. A complete classification for fermionic SPT (FSPT) phases can also be obtained by the so-called general group super-cohomology theory~\cite{special,general1,general2}, spin cobordism theory~\cite{Kapustin2014,cobordism}, or by gauging the corresponding global symmetry~\cite{Gu-Levin,gauging1,gauging3,dimensionalreduction,gauging2,2DFSPT,braiding}.

Very recently, crystalline SPT phases have been intensively studied for free-fermion and interacting bosonic systems~\cite{TCI,ITCI,reduction,building,correspondence,SET,BCSPT,Shiozaki2018,ZDSong2018,defect,realspace,KenX,rotation,LuX}. 
These states are not only of conceptual importance, but also provide great opportunities for experimental realization~\cite{TCIrealization1,TCIrealization2,TCIrealization3,TCIrealization4}. In particular, an explicit block-state construction scheme for crystalline SPT phases was established in Ref.~\onlinecite{reduction}. Furthermore, it has been highlighted that the classification of space group SPT phases is closely related to SPT phases with internal symmetry. In Ref.~\onlinecite{correspondence}, a ``crystalline equivalence principle'' was proposed: i.e., crystalline topological phases with symmetry $G$ are in one-to-one correspondence with topological phases protected by the same internal symmetry $G$, but acting in a twisted way, where if an element of $G$ is a mirror reflection, it should be regarded as a time-reversal symmetry.
Thus, the classification of crystalline SPT phases for interacting bosonic and free-fermion systems can be computed systematically.
For interacting fermion systems, the strategy of classification schemes has also been discussed~\cite{correspondence,defect,realspace,KenX} and some simple examples have been studied~\cite{rotation,YMLu2018}; however, a detailed understanding of generic cases is still lacking.

In this paper we systematically study the construction and classification of 2D FSPT phases protected by point group symmetry via a block-state decoration scheme. In particular,  we discover several intriguing fermionic point group SPT phases that cannot be realized in either free-fermion or in interacting boson systems. Table \ref{classification} summarizes the classification results. We also compare these results with the classification of 2D FSPT phases with the corresponding internal symmetry.
We conjecture a fermionic crystalline equivalence principle, stating that a mirror reflection symmetry action should be mapped onto a time reversal symmetry action, and that spinless(spin-1/2) fermion systems should be mapped onto spin-1/2(spinless) fermion systems. The possibility of experimental realization is also addressed.

\begin{table}[t]
\renewcommand\arraystretch{1.2}
\begin{tabular}{|c|c|c|c|c|}
\hline
\diagbox{$~G_b$}{spin}&$~\text{spinless}~$&$~\text{spin-1/2}~$\\
\hline
$C_{2m-1}$&$\mathbb{Z}_{2m-1}$&$~\mathbb{Z}_{2m-1}~$\\
\hline
$C_{2m}$&$\mathbb{Z}_m$&$\left\{\begin{aligned}&\mathbb{Z}_2\times\mathbb{Z}_{4m},~&m\in\mathrm{even}\\
&\mathbb{Z}_{8m},~&m\in\mathrm{odd}\end{aligned}\right.$\\
\hline
$D_{2m-1}$&$\mathbb{Z}_2$&$~\mathbb{Z}_1~$\\
\hline
$D_{2m}$&$\mathbb{Z}_2$&$~\mathbb{Z}_2\times\mathbb{Z}_2~$\\
\hline
\end{tabular}
\caption{The classification of interacting 2D FSPT phases with point group symmetry for spinless fermions and spin-1/2 fermions.}
\label{classification}
\end{table}

\textit{A simple example with a $D_4$ symmetry} -- It is well known that there are 10 point groups in 2D, classified as cyclic groups $C_n$ and dihedral groups $D_n$ ($n=1,2,3,4,6$). As the $C_n$ cases have already been discussed in Ref. \onlinecite{rotation}, here we mainly focus on the $D_n$ cases. Mathematically, each dihedral group is a semiproduct of a rotation and a reflection symmetry group $D_n=C_n\rtimes\mathbb{Z}_2^{\mathrm{M}}$. It eventuates that the most interesting and complicated cases arise for even numbers of $n$. Below, we will begin with the most intriguing case for spinless fermion systems, namely, the case protected by a $D_4$ symmetry (the simplest non-Abelian point group with even number $n$). 

Similar to Ref. \onlinecite{reduction}, we begin with the ``extended trivialization'' of $D_4$. Suppose a $|\psi\rangle$ is an SPT state that cannot be trivialized by a symmetric finite-depth local unitary transformations,
we can still act with an alternative local unitary to \textit{extensively} trivialize $|\psi\rangle$. First, we trivialize the region $U$ (see Fig. \ref{D4 trivialization}): i.e., restrict $O^{\mathrm{loc}}$ to $U$ as $O^{\mathrm{loc}}_U$ and act it on $|\psi\rangle$:
\begin{align}
O^{\mathrm{loc}}_U|\psi\rangle=|T_U\rangle\otimes|\psi_{\bar{U}}\rangle
\end{align}
where the system is in the product state $|T_U\rangle$ in region $U$ and the remainder of the system $\bar{U}$ is in the state $|\psi_{\bar{U}}\rangle$. To trivialize the system symmetrically, we denote that $V_gO^{\mathrm{loc}}_UV_g^{-1}$ trivializes $gU$ ($g\in D_4$, see Fig. \ref{D4 trivialization}). Therefore, we act on $|\psi\rangle$ with:
\begin{align}
O^{\mathrm{loc}}_R=\bigotimes\limits_{g\in D_4}V_gO^{\mathrm{loc}}_UV_g^{-1}
\end{align}
which results in:
\begin{align}
|\psi'\rangle=O^{\mathrm{loc}}_R|\psi\rangle=\bigotimes\limits_{g\in D_4}|T_{gU}\rangle\otimes\bigotimes\limits_{i=1,a}^{4,d}|\psi_i\rangle\otimes|\psi_{\mathrm{0D}}\rangle
\label{extensive trivialization}
\end{align}
Now, all nontrivial properties of $|\psi\rangle$ are encoded in lower-dimensional block-states $|\psi_i\rangle$ and $|\psi_{\mathrm{0D}}\rangle$.

Next, we consider the 1D block-state $|\psi_{i}\rangle$. Let us divide 1D blocks into two categories: i.e., category-$\1$ $(1\sim4)$ and category-$\2$ $(a\sim d)$. As these two categories are independent under $D_4$ symmetry, we can discuss the decorations on these categories separately. We investigate the decoration on category-$\1$ ($\mathbb{Z}_2^f$ is the fermion parity, $\bs{R}$ and $\bs{M}$ are rotation and reflection generators of $D_4$). Decorations on $(1,3)/(2,4)$: $(1,3)/(2,4)$ are invariant under $\bs{M}/\bs{MR}^2$, and the block-state should consist of 1D FSPT phases with $\mathbb{Z}_2^f\times\mathbb{Z}_2$ internal symmetry, and be compatible with all other space group symmetries.

The classification of 1D-invertible topological order (ITO) with $\mathbb{Z}_2^f\times\mathbb{Z}_2$ symmetry for interacting fermion systems is $\mathbb{Z}_2^2$ described by the following root phases \cite{general2}:
\begin{enumerate}
\item The Kitaev Majorana chain \cite{Majorana,1Dfermion};
\item An FSPT state that
can be realized by two Majorana chains \cite{special,general1,general2}.
\end{enumerate}
\begin{figure}
\centering
\includegraphics[width=0.35\textwidth]{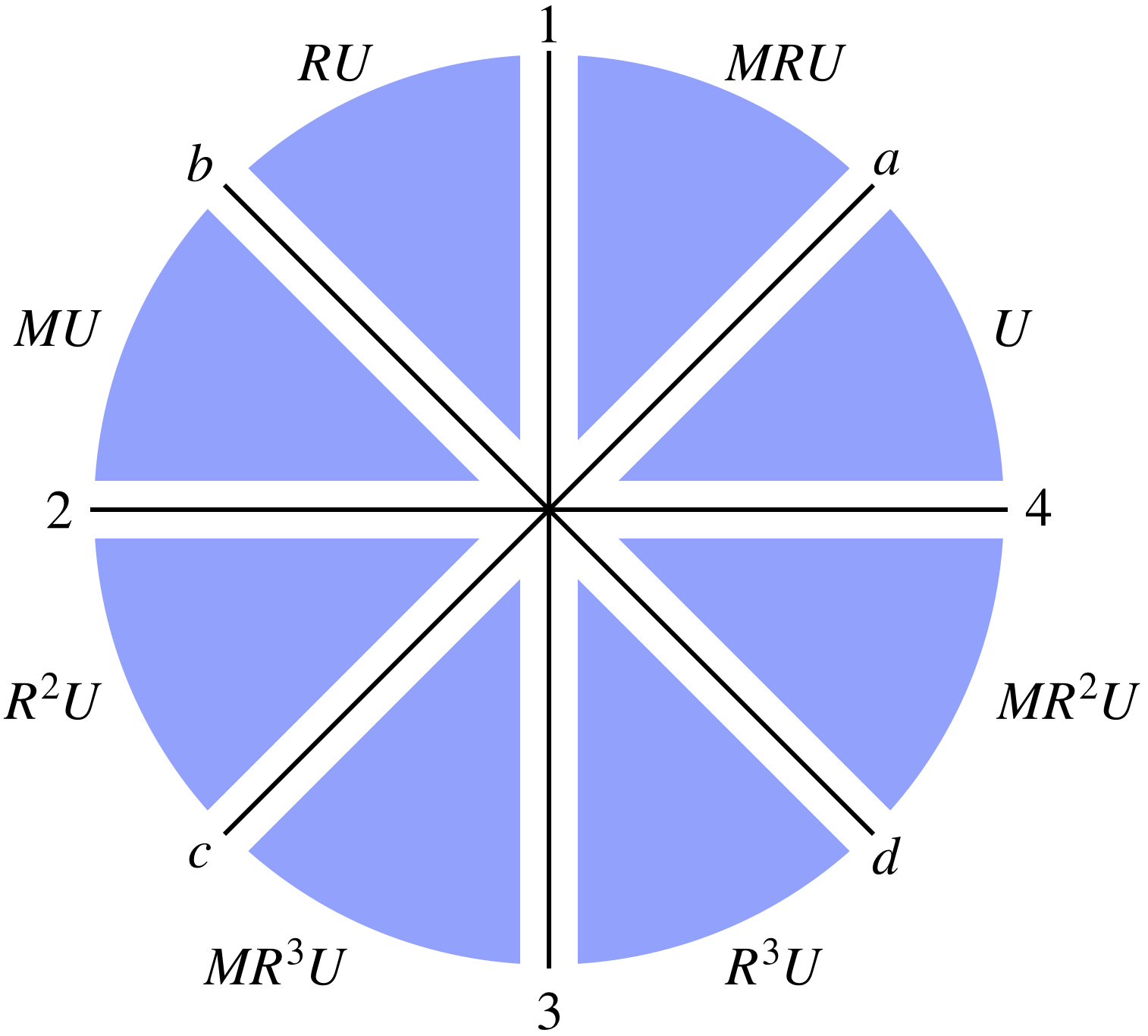}
\caption{Extended trivialization of 2D FSPT phases with dihedral group $D_4$. Here all shadowed regimes are trivialized according to Eq. (\ref{extensive trivialization}).}
\label{D4 trivialization}
\end{figure}

We note that all of these can be realized by fixed-point wavefunctions and exact-soluble parent Hamiltonians. First, we investigate the Majorana chain decoration. Considering four Majorana chains decorated on category-$\1$, there are four Majorana modes $(\gamma_1,\gamma_2,\gamma_3,\gamma_4)$ located at the 0D block, with the local fermion parity symmetry $P_f=-\gamma_1\gamma_2\gamma_3\gamma_4$ which is odd under rotation:
$\bs{R}P_f\bs{R}^{-1}=-\gamma_2\gamma_3\gamma_4\gamma_1=-P_f$.
Hence, these four Majorana modes form a projective representation of the group $C_4\times\mathbb{Z}_2^f$ as a subgroup of $D_4\times\mathbb{Z}_2^f$. Therefore, a non-degenerate ground state is forbidden. As a consequence, Majorana chain decoration on category-$\1$ is forbidden by $D_4$ symmetry, and the above argument is also applicable for category-$\2$. Note that if we consider all 1D blocks together and decorate a Majorana chain on each, $P_f$ commutes with rotation. Nevertheless, it is simple to verify that $P_f$ anti-commutes with  reflection, thus Majorana chain decoration remains forbidden by $D_4$.

Subsequently, we investigate the decoration of 1D FSPT states on category-$\1$. Consider the geometry shown in Fig. \ref{D4 center}, with eight Majorana modes located at the rotation center: $(\gamma_1,\gamma_2,\gamma_3,\gamma_4)$ and $(\gamma_1',\gamma_2',\gamma_3',\gamma_4')$, with rotation and reflection symmetry ${\bs{R}}^4=1$ and ${\bs{M}}^2=1$:
\begin{align}
\bs{R}:~\gamma_i\mapsto\gamma_{i+1},~~\gamma_i'\mapsto\gamma_{i+1}',~~i=1,2,3,4
\label{symmetry R}
\end{align}
\begin{align}
\bs{M}:~\gamma_i\mapsto-\gamma_{6-i},~~\gamma_i'\mapsto\gamma_{6-i}',~~i=1,2,3,4
\label{symmetry M}
\end{align}
All subscripts take the values with modulo 4, e.g., $\gamma_{5}\equiv\gamma_{1}$ and $\gamma_{5}^\prime\equiv\gamma_{1}^\prime$. Now we explain why we need to introduce the above symmetry properties for Majorana modes. Since the rotation properties are easier to understand, we mainly focus on the reflection properties below. For the vertical axis, $\bs{M}$ acts as an on-site $\mathbb{Z}_2$ symmetry, and the two Majorana modes at the edge of the decorated 1D FSPT state should anti-commute with the fermion parity. Thus, for $i=1,3$, $\bs{M}:\gamma_i\mapsto-\gamma_i,~\gamma_i'\mapsto\gamma_i'$.
Similarly, for the horizontal axis, 
$\bs{MR}^2$ acts as an on-site $\mathbb{Z}_2$ symmetry, so for $j=2,4$, $\bs{MR}^2:\gamma_j\mapsto-\gamma_j,~\gamma_j'\mapsto\gamma_j'$. Together with rotational symmetry Eq. (\ref{symmetry R}), it is easy to verify that for $j=2,4$,
$\bs{M}:\gamma_j\mapsto-\gamma_{6-j},~\gamma_j'\mapsto\gamma_{6-j}'$.

\begin{figure}
\centering
\includegraphics[width=0.3\textwidth]{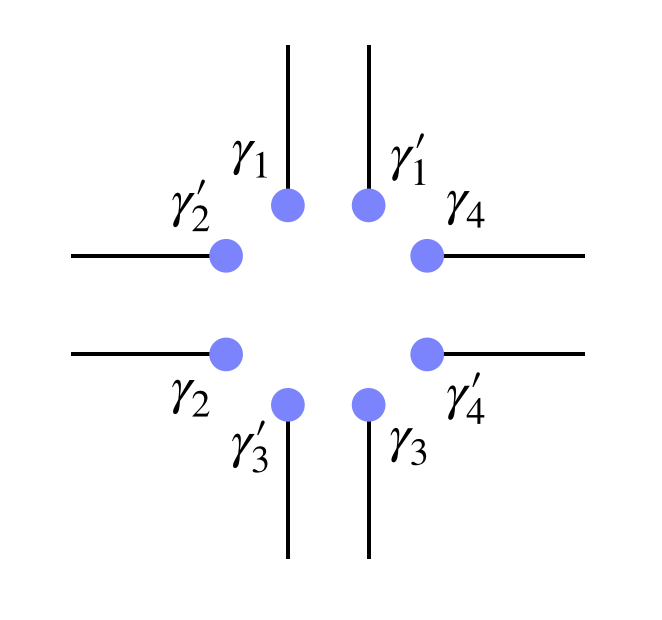}
\caption{0D block of 2D FSPT with $D_4$ symmetry, corresponding to 1D FSPT phase decoration. Blue dots represent the Majorana zero modes on the edge of decorated root phases.}
\label{D4 center}
\end{figure}

Finally, we try to gap out these Majorana modes through interactions in a symmetric way. First, we consider the following interacting Hamiltonian:
\begin{align}
\mathcal{H}_{U}=U\big[\gamma_1\gamma_1'\gamma_3\gamma_3'+\gamma_2\gamma_2'\gamma_4\gamma_4'\big], \quad U>0
\label{D4 gap}
\end{align}
For the ground state, 
\begin{align}\gamma_1\gamma_1'\gamma_3\gamma_3'=\gamma_2\gamma_2'\gamma_4\gamma_4'=-1.
\label{constraint}
\end{align}
The ground state is four-fold degenerate from Eq. (\ref{D4 gap}). To lift this degeneracy, we can further add a term:
\begin{align}
\mathcal{H}_{J}=J(\gamma_1\gamma_2\gamma_1'\gamma_2'+\gamma_1\gamma_2\gamma_3'\gamma_4'), \quad J>0
\label{gapped}
\end{align}
Consider the total Hamiltonian $\mathcal{H}=\mathcal{H}_{U}+\mathcal{H}_{J}$ and take the limit $U\rightarrow\infty$, such that it leads to the constraint Eq. (\ref{constraint}). Within the constraint subspace, Hamiltonian (\ref{gapped}) is symmetric under $D_4$ symmetry. Then, because both terms in $\mathcal{H}_{J}$ commute with each other and have eigenvalues $\pm J$, $\mathcal{H}_{J}$ has a unique ground state with eigenvalue $-2J$. Thus, we can lift the degeneracy in a $D_4$ symmetric way and this decoration is compatible with $D_4$ symmetry.

Below, we argue that such a 1D block-state decoration cannot be trivialized. Considering a 2D system with an open boundary (see Fig. \ref{D4 boundary}), we further place four additional Majorana chains $(\alpha,\beta,\gamma,\delta)$ on the boundary, adding eight additional more Majorana modes $(\gamma_j,\gamma_j'),~j=\alpha,\beta,\gamma,\delta$. For any group of four Majorana modes, e.g., $(\gamma_1,\gamma_1',\gamma_\alpha,\gamma_\delta')$, at one side of the edge with the following reflection symmetry properties:
\begin{align}
\bs{M}:~(\gamma_1,\gamma_1',\gamma_\alpha,\gamma_\delta')\mapsto(-\gamma_1,\gamma_1',\gamma_\delta',\gamma_\alpha).
\end{align}
this group will be gapped without breaking the reflection symmetry due to the compatibility of local fermion parity  $P_f=-\gamma_1\gamma_1'\gamma_\alpha\gamma_\delta'$ \cite{supplementary}.
\begin{figure}
\centering
\includegraphics[width=0.296\textwidth]{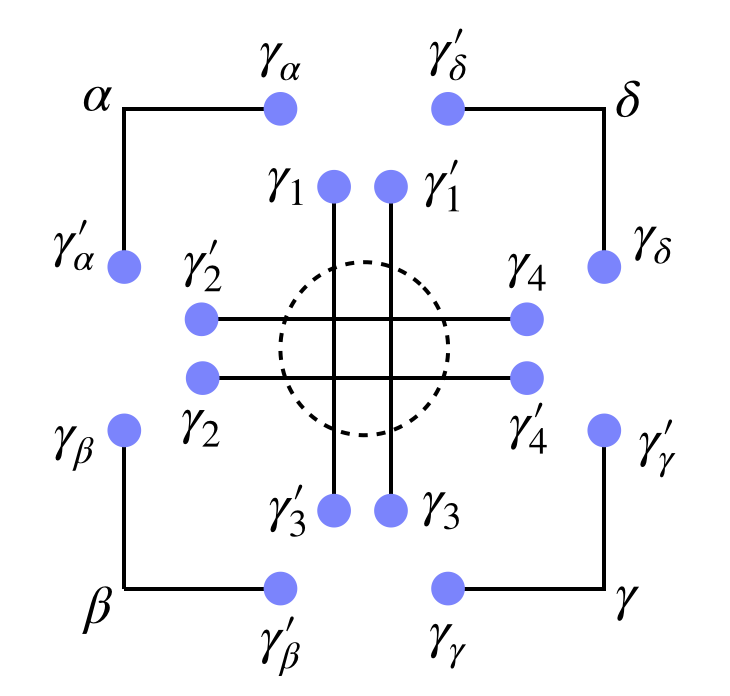}
\caption{Forbidden trivialization of 1D FSPT phase-decoration on category-$\1$ for spinless fermions. Again, blue dots represent the Majorana zero modes.} 
\label{D4 boundary}
\end{figure}
Nevertheless, it is remains unclear whether such a ``trivialization scheme” is compatible with the full $D_4$ symmetry. Considering the Majorana chain labeled by $\alpha$, the symmetry $\bs{M'}\equiv\bs{MR}^3\in D_4$ acts on $\alpha$ as an effective reflection symmetry. However, because a single open Majorana chain is incompatible with reflection symmetry ${\bs{M'}}^2=1$(anti-commutes with the total fermion parity), 
this suggests that the boundary Majorana modes cannot be gapped out without breaking the full $D_4$ symmetry. As a result, we conclude that the 1D FSPT state decoration on category-$\1$ must describe a non-trivial 2D FSPT state with $D_4$ symmetry. In particular, the 2D FSPT state that we have constructed here is an intrinsic interacting FSPT state that cannot be realized by free-fermion systems \cite{noninteracting,TBT1,TBT2,TBT3} or interacting bosonic systems.

Similar arguments hold for category-$\2$. Thus, the question naturally arises of whether these two cases describe independent FSPT states or not. Let us consider the geometry in Fig. \ref{D4 stack}, where we place eight different Majorana chains labeled by $\alpha\sim\theta$ on the boundary of the system. It is simple to verify that this assignment respects $D_4$ symmetry. Then, as aforementioned, each Majorana mode on the boundary can be symmetrically gapped out without breaking the two reflection symmetries $\bs{M}$ and $\bs{M'}$. 
Therefore, this case indeed corresponds to a trivial bulk because a gapped, short-range entangled symmetric boundary termination is obtained. That is, the cases of decorating 1D FSPT phases on category-$\1$ and category-$\2$ are non-independent and these two types of decorations give rise to only one nontrivial FSPT state.

\begin{figure}
\centering
\includegraphics[width=0.35\textwidth]{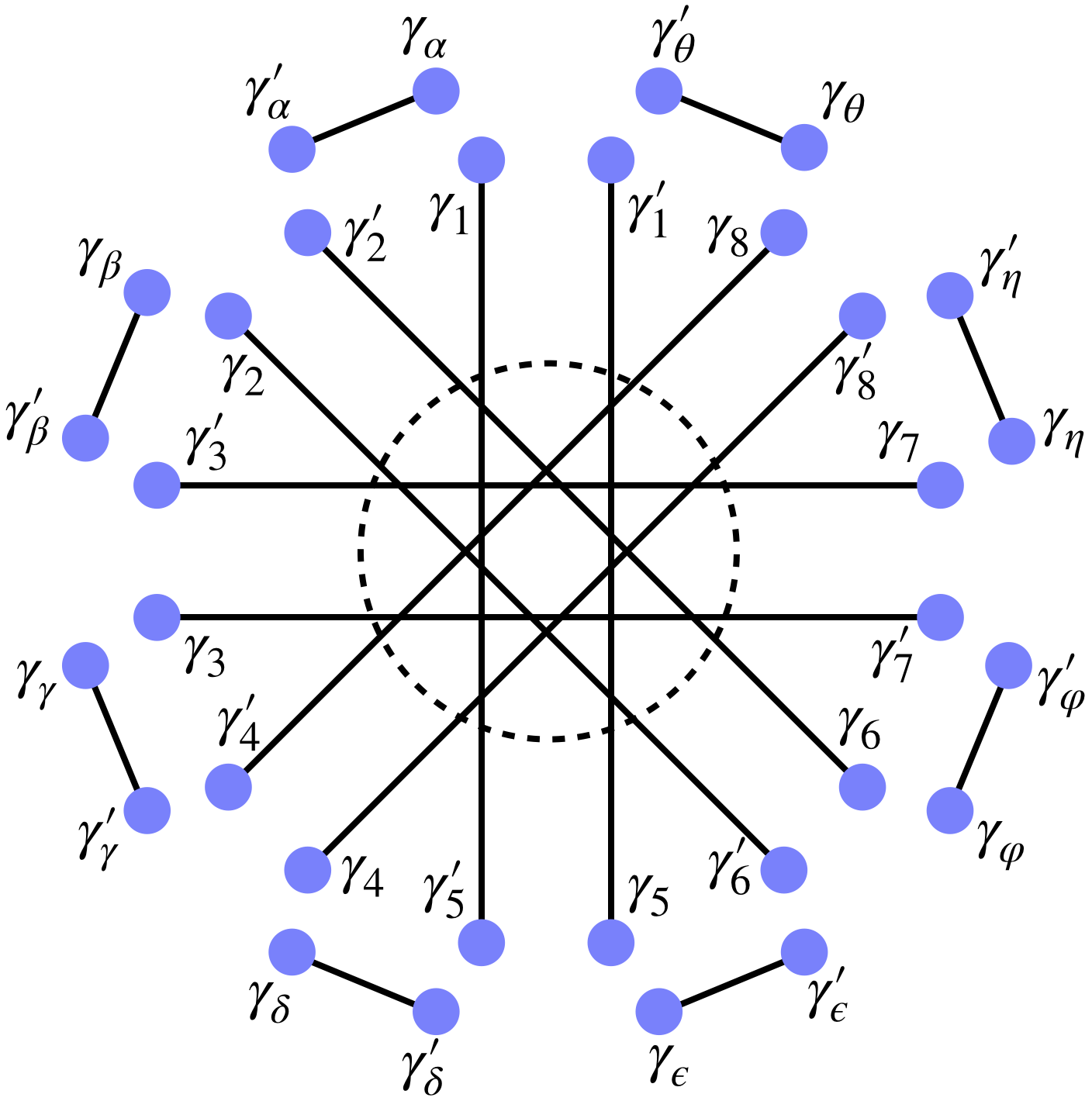}
\caption{Trivialization of decorated 1D FSPT state on both category-I and category-II 1D blocks for spinless fermion systems.}
\label{D4 stack}
\end{figure}

Now we consider the 0D block-state $|\psi_{\mathrm{0D}}\rangle$. As $D_4$ acts on the 0D block as an internal symmetry, the full data of the 0D block-state are \cite{general2}:
\begin{align}
\mathcal{H}^0\big(\mathbb{Z}_4\rtimes\mathbb{Z}_2,\mathbb{Z}_2\big)\times\mathcal{H}^1\big[\mathbb{Z}_4\rtimes\mathbb{Z}_2,U(1)\big]=\mathbb{Z}_2\times\mathbb{Z}_2^2
\label{classification data}
\end{align}
We first consider an atomically insulating state with four complex fermions 
\begin{align}
|\phi\rangle_{\mathrm{0D}}=c_1^\dag c_2^\dag c_3^\dag c_4^\dag|0\rangle,
\label{D4 atomic}
\end{align}
with the following symmetry properties (all subscripts take the value of modulo 4):
\begin{align}
\bs{R}:~c_i^\dag\mapsto c_{i+1}^\dag,~~\bs{M}:~c_i^\dag\mapsto c_
{6-i}^\dag,~~i=1,2,3,4
\label{symmetry 0D}
\end{align}
Again, all subscripts take values of modulo 4, and the above symmetry actions on Eq. (\ref{D4 atomic}) give rise to:
\begin{align}
\bs{R}|\psi\rangle_{\mathrm{0D}}=\bs{M}|\psi\rangle_{\mathrm{0D}}=-|\psi\rangle_{\mathrm{0D}}
\label{0D trivialization}
\end{align}
Thus, the eigenvalue $-1$ of rotation symmetry and reflection symmetry indeed corresponds to a topological trivial state. In Ref. \onlinecite{rotation}, a closed Majorana chain surrounding the 0D block is introduced to trivialize the 0D block-state with odd fermion parity. This construction can also be applied here
the 0D block-state with odd fermion parity will also be trivialized (see Supplementary Material \cite{supplementary} for full details). Therefore, all 0D block-states are trivialized and the classification of 2D FSPT phases protected by dihedral symmetry $D_4$ for spinless fermions is $\mathbb{Z}_2$ (see Table \ref{classification}). This classification coincides with the classification of the 2D FSPT protected by internal symmetry $\mathbb{Z}_4\rtimes\mathbb{Z}_2^{\mathrm{T}}$(where 
$\mathbb{Z}_2^{\mathrm{T}}$ is time-reversal symmetry) for spin-$1/2$ fermions (see Table \ref{brute force classification}).

\begin{table}[t]
\centering
\renewcommand\arraystretch{1.2}
\begin{tabular}{|c|c|c|c|c|}
\hline
\diagbox{$~G_b$}{spin}&~\text{spinless}~&~spin-$1/2$~~\\
\hline
$~\mathbb{Z}_{2m-1}~$&$~\mathbb{Z}_{2m-1}~$&$\mathbb{Z}_{2m-1}$\\
\hline
$~\mathbb{Z}_{2m}~$&$\left\{\begin{aligned}&\mathbb{Z}_2\times\mathbb{Z}_{4m},~&m\in\mathrm{even}\\
&\mathbb{Z}_{8m},~&m\in\mathrm{odd}\end{aligned}\right.$&$\mathbb{Z}_m$\\
\hline
$~\mathbb{Z}_{2m-1}\rtimes\mathbb{Z}_2^{\mathrm{T}}~$&$~\mathbb{Z}_1~$&$\mathbb{Z}_2$\\
\hline
$~\mathbb{Z}_{2m}\rtimes\mathbb{Z}_2^{\mathrm{T}}~$&$~\mathbb{Z}_2\times\mathbb{Z}_2~$&$\mathbb{Z}_2$\\
\hline
\end{tabular}
\caption{The interacting classification of 2D FSPT phases with internal symmetries,  for spinless and spin-$1/2$ fermions, respectively. $\mathbb{Z}_2^{\mathrm{T}}$ is time-reversal symmetry.} 
\label{brute force classification}
\end{table}

Finally, we discuss systems with spin-$1/2$ fermions. Through similar block-state constructions, we obtain a $\mathbb{Z}_2^2$ classification: 1D block-state decorations do not contribute to any nontrivial FSPT phase, because for 1D systems with spin-$1/2$ fermions and $\mathbb{Z}_2$ symmetry (total symmetry group is $\mathbb{Z}_4^f$), there is no nontrivial SPT phase \cite{general2}. For the 0D block, the first $\mathbb{Z}_2$ of Eq. (\ref{classification data}) is not allowed \cite{general2}, and Eq. (\ref{0D trivialization}) has no nontrivial eigenvalue under rotation and reflection.(We note that for spin-1/2 fermions, there would be an extra $i$ factor in Eq. (\ref{symmetry 0D}) with $\bs{R}^2=\bs{M}^2=-1$, which cancels the $-1$ in Eq. (\ref{0D trivialization}).) As a result, there is no trivialization in this case.  
Thus, 0D block-state decorations contribute to a $\mathbb{Z}_2^2$ classification and the overall classification of 2D FSPT phases with $D_4$ symmetry for spin-$1/2$ fermions is $\mathbb{Z}_2^2$.

\textit{Point group FSPT with general $D_n$ symmetry} -- Spinless and spin-1/2 fermion systems with $D_2$ and $D_6$ point group symmetry can also be constructed in a similar way and the classification results are exactly the same as the $D_4$ case.
Moreover, a 2D FSPT state protected by $D_n$ symmetry with odd $n$ can also be constructed by similar block-state decorations as aforementioned. 
In fact, the essential contributions of nontrivial FSPT phases in these cases solely derive from the reflection subgroup. All results of classification are summarized in Table \ref{classification}, and the full details of the wavefunction constructions can be found in Supplementary Material \cite{supplementary}.


\textit{Generalized crystalline equivalence principle} -- To this end, we would like to examine the generalized crystalline equivalence principle for 2D interacting fermion systems:  
we calculate the classification of 2D FSPT phases with internal symmetry $\mathbb{Z}_n$ and $\mathbb{Z}_n\rtimes\mathbb{Z}_2^{\mathrm{T}}$ using the so-called general group super-cohomology theory\cite{general2}. The classification results are shown in Table \ref{brute force classification} (see Supplementary Material \cite{supplementary} for detailed calculations). Comparing the results in Table \ref{classification} and \ref{brute force classification}, we conjecture that the crystalline equivalence principle can be generalized into 2D interacting fermion systems. w
We should map the \textit{mirror reflection symmetry} onto an internal \textit{time-reversal symmetry} and we should also map spinless(spin-1/2) fermions onto spin-1/2(spinless) fermions. The twist on spinless and spin-$1/2$ fermions can be naturally interpreted as the spin rotation of fermions: a $2\pi$ rotation of a fermion around a specific axis results in a $-1$ phase factor (see Supplementary Material \cite{supplementary} for more details).

\textit{Conclusion and discussion} -- In this paper, we systematically constructed and classified the 2D interacting FSPT phases with point group symmetry using explicit block-state constructions (with $D_4$ symmetry as a concrete example). 
Our results also verify the generalized crystalline equivalence principle for 2D interacting fermionic systems. Experimentally, single-layer iron selenide (FeSe, an iron-based superconductor with space group $P4/nmm$ in 3D \cite{iron, selenide1,selenide2,selenide3}) on a ferromagnetic substrate could be a natural candidate for realizing an intrinsic FSPT phase with $D_4$ symmetry \cite{private}. (We note that the fermions in iron selenide are spin-polarized due to the ferromagnetic proximity, and can thus be effectively treated as spinless.) According to our block-state construction, there are several dangling Majorana fermions located at the boundary of the system, and these gapless modes can be detected by experiments. The boundary Majorana modes via block-state construction may be related to the recent proposed ``corner Majorana modes" in monolayer iron selenides \cite{RXZhang2019,CXLiu2019}. Furthermore, the bulk topological defects which can be detected by STM tomography can also be the evidences of the designed SPT states.
Finally, we stress that the method proposed here is also applicable to space group symmetry, and that exploration of its higher-dimensional generalizations would be a very interesting future direction.

\begin{acknowledgements}
\textit{Acknowledgement} -- We thank Jianjian Miao, Liujun Zou, Chenjie Wang and Xie Chen for stimulating discussions, and especially Wei Li for sharing his unpublished experimental data. This work is supported by Direct Grant No. 4053346 from The Chinese University of Hong Kong and funding from Hong Kong's Research Grants Council (GRF No.14306918, ANR/RGC Joint Research Scheme No. A-CUHK402/18). SY is supported by NSFC (Grant No. 11804181) and the National Key R\&D Program of China (Grant No. 2018YFA0306504).
\end{acknowledgements}

\providecommand{\noopsort}[1]{}\providecommand{\singleletter}[1]{#1}%
%




\pagebreak

\clearpage

\appendix
\setcounter{equation}{0}
\newpage

\renewcommand{\thesection}{S-\arabic{section}} \renewcommand{\theequation}{S%
\arabic{equation}} \setcounter{equation}{0} \renewcommand{\thefigure}{S%
\arabic{figure}} \setcounter{figure}{0}

\centerline{\textbf{Supplemental Material}}

\maketitle

\section{\1. Twist of spinless/spin-$1/2$ fermions\label{math}}
In the main text we conjectured a generalized ``crystalline equivalence principle'' for 2D interacting fermionic systems, with a twist on spinless and spin-$1/2$ fermions. In this section we discuss about the physical understanding of this peculiar twist. 
\subsection{A. Central extension of fermion parity}
Different extensions of fermion parity (labeled by $\mathbb{Z}_2^f$) of the symmetry group characterize the systems with spinless and spin-$1/2$ fermions. For a interacting fermionic system with a physical symmetry group $G_b$, the total symmetry group $G_f$ is a central extension of fermion parity $\mathbb{Z}_2^f$:
\begin{align}
0\lra\mathbb{Z}_2^f\lra G_f\lra G_b\lra0
\label{extension of fermion parity}
\end{align}
and different central extensions are characterized by different factor systems of Eq. (\ref{extension of fermion parity}). 

\textbf{\underline{Lemma} (Factor System)} For a group $(G,\cdot)$, an Abelian group $(A,+)$, and a short exact sequence:
\begin{equation}
0\lra A\lra X\lra G\lra0
\label{AXG}
\end{equation}
There is a factor system of the short exact sequence Eq. (\ref{AXG}) who consists the function $f$ and a homomorphism $\sigma$:
\begin{align}
\left.
\begin{aligned}
f:G\times G~&\lra~~~A\\
(g,h)~&\longmapsto f(g,h)
\end{aligned}
\right.~,~~\left.
\begin{aligned}
\sigma:G~~&\lra \text{End}(A)\\
g~~&\longmapsto~~~\sigma_g
\end{aligned}
\right.
\end{align}
such that it makes the Cartesian product $G\times A$ a group $X$ with multiplication: \begin{align}
(g,a)*(h,b)=(g\cdot h,f(g,h)+a+\sigma_g(b))
\end{align}
And $f$ must be a group 2-cocycle which is classified by 2 group cohomology: $f\in\mathcal{H}^2(G,A)$. Where End$(A)$ is the endomorphism of group $A$.

\begin{figure}[t]
\centering
\includegraphics[width=0.5\textwidth]{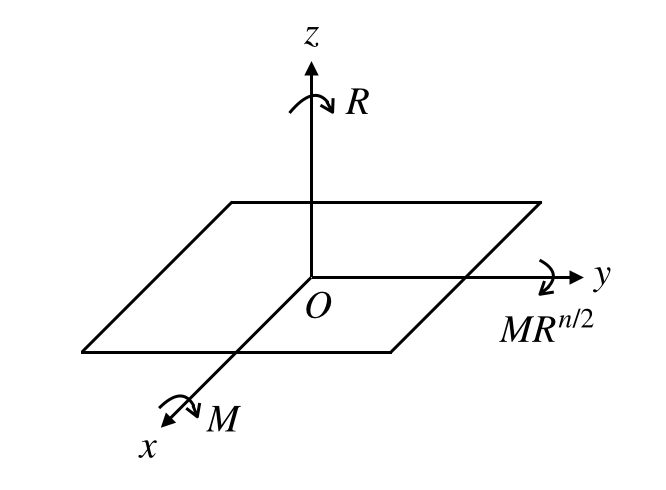}
\caption{Rotations of topological spin around 3 axes, generated by $\bs{M}^2$, $(\bs{MR}^{n/2})^2$ and $\bs{R}^n$ symmetry operations, respectively.}
\label{fermion parity twist}
\end{figure}

For even-fold dihedral group, different extensions of fermion parity [cf. Eq. (\ref{extension of fermion parity})] are characterized by the $\mathbb{Z}_2$-valued group 2-cohomology of even-order dihedral group $D_n$ is:
\begin{align}
\mathcal{H}^2(D_n,\mathbb{Z}_2)=\mathbb{Z}_2^3
\end{align}
which are labeled by $\bs{R}^{n},\bs{M}^2,(\bs{MR}^{n/2})^2=\pm1$ ($\bs{R}$ and $\bs{M}$ are rotation and reflection generators of dihedral group). More precisely, they are labeled by the following 2-cocycles:
\begin{align}
\left\{
\begin{aligned}
&\omega_2(\bs{R}^{n/2},\bs{R}^{n/2})=\pm1\\
&\omega_2(\bs{M},\bs{M})=\pm1\\
&\omega_2(\bs{MR}^{n/2},\bs{MR}^{n/2})=\pm1
\end{aligned}
\right.
\label{2-cocycle}
\end{align}
It is easy to verify that 2-cocycles in Eq. (\ref{2-cocycle}) are independent with each other and gauge invariant [i.e., Eq. (\ref{2-cocycle}) is invariant if we add arbitrary 2-coboundary to each of them]. 

\subsection{B. Physical understanding of twist}
The most physical relevant cases are systems with spinless or spin-$1/2$ fermions, which are characterized by the following abbreviated 2-cocycles:
\[
\bs{R}^{n}=\bs{M}^2=(\bs{MR}^{n/2})^2=\pm1
\]
where $+/-$ corresponds to spinless/spin-$1/2$ fermions. In this subsection we discuss the physical understanding of these 2-cocycles.

$n$-fold rotational operation $\bs{R}^n$ generates a $2\pi$ rotation of topological spin of fermions around $z$-axis; 2-fold reflection operation $\bs{M}^2$ generates a $2\pi$ rotation of topological spin of fermions around $x$-axis; 2-fold operation of a composition of $n/2$-fold rotational operation and reflection operation (effectively, this composite is the reflection operation with respect to $y$-axis), $(\bs{M}\bs{R}^{n/2})^2$, generates a $2\pi$ rotation of topological spin of fermions around $y$-axis (cf. Fig. \ref{fermion parity twist}).  All of aforementioned $2\pi$ rotations of topological spin of fermions will result a $-1$ phase factor of the wavefunction of fermions. As the consequence, the generalized ``crystalline equivalence principle'' in (2+1)D interacting fermionic systems should have a twist of fermion parity extensions due to the aforementioned $2\pi$ rotation of topological spin of fermions around 3 axes, respectively, which are labeled by 2-cocycles $\omega_2\in\mathcal{H}^2(D_n,\mathbb{Z}_2)$, as shown in Eq. (\ref{2-cocycle}). Equivalently, the twist of spinless and spin-$1/2$ fermions can be understood by rotations of topological spins of fermions.

\section{\2. Symmetry property of the gapping Hamiltonian}
In the main text, we discussed the FSPT phases protected by $D_4$ symmetry for spinless fermions, and the unique nontrivial SPT state can be obtained by 1D FSPT state decoration. In particular, we use 2 interacting Hamiltonians [Eqs. (\ref{D4 gap}) and Eq. (\ref{gapped}) in the main text] in order to gap out the dangling Majorana fermions as the edge modes of the decorated 1D FSPT states near the 0D block. We investigate the symmetry properties of these 2 interacting Hamiltonians in this section. 

The symmetry properties of Majorana fermions are summarized as Eq. (\ref{symmetry R}) in the main text, thus we can discuss the symmetry properties of Hamiltonian directly. Under rotation and reflection, 
\begin{align}
\left.
\begin{aligned}
\bs{R}:~&\gamma_1\gamma_1'\gamma_3\gamma_3'\mapsto\gamma_2\gamma_2'\gamma_4\gamma_4'\\
&\gamma_2\gamma_2'\gamma_4\gamma_4'\mapsto\gamma_3\gamma_3'\gamma_1\gamma_1'=\gamma_1\gamma_1'\gamma_3\gamma_3'\\
\bs{M}:~&\gamma_1\gamma_1'\gamma_3\gamma_3'\mapsto(-\gamma_1)\gamma_1'(-\gamma_3)\gamma_3'=\gamma_1\gamma_1'\gamma_3\gamma_3'\\
&\gamma_2\gamma_2'\gamma_4\gamma_4'\mapsto(-\gamma_4)\gamma_4'(-\gamma_2)\gamma_2'=\gamma_2\gamma_2'\gamma_4\gamma_4'
\end{aligned}
\right.
\end{align}
Hence $\mathcal{H}_U$ is invariant under $D_4$ symmetry. In large $U$ limit, $\mathcal{H}_U$ gives the constraint as Eq. ({\color{red}7}) in the main text, which can be rephrased as following:
\begin{align}
\gamma_1\gamma_3'=\gamma_3\gamma_1',~~\gamma_2\gamma_4'=\gamma_4\gamma_2'
\end{align}
Therefore, under rotation and reflection,
\begin{align}
\bs{R}:~\gamma_1\gamma_1'\gamma_2\gamma_2'\mapsto\gamma_2\gamma_2'\gamma_3\gamma_3'=\gamma_1\gamma_1'\gamma_2\gamma_2'
\end{align}
\begin{align}
\bs{R}:~\gamma_1\gamma_2\gamma_3'\gamma_4'\mapsto &\gamma_2\gamma_3\gamma_4'\gamma_1'=-\gamma_2\gamma_3\gamma_1'\gamma_4'\nonumber\\
=&-\gamma_2\gamma_1\gamma_3'\gamma_4'=\gamma_1\gamma_2\gamma_3'\gamma_4'
\end{align}
\begin{align}
\bs{M}:~\gamma_1\gamma_1'\gamma_2\gamma_2'\mapsto &(-\gamma_1)\gamma_1'(-\gamma_4)\gamma_4'=\gamma_1\gamma_1'\gamma_4\gamma_4'\nonumber\\
=&\gamma_1\gamma_1'\gamma_2\gamma_2'
\end{align}
\begin{align}
\bs{M}:~\gamma_1\gamma_2\gamma_3'\gamma_4'\mapsto &(-\gamma_1)(-\gamma_4)\gamma_3'\gamma_2'=-\gamma_1\gamma_3'\gamma_4\gamma_2'\nonumber\\
=&-\gamma_1\gamma_3'\gamma_2\gamma_4'=\gamma_1\gamma_2\gamma_3'\gamma_4'
\end{align}
Hence $\mathcal{H}_J$ is invariant under $D_4$ symmetry. Therefore, the interacting Hamiltonians we used to gap out the dangling Majorana fermions as the edge modes of the decorated 1D FSPT states near the 0D block are $D_4$ symmetric.

The symmetry properties of interacting Hamiltonian can be manifested more clearly and transparent by an alternative notion. Define 4 complex fermions:
\begin{align}
\left.
\begin{aligned}
&c_1^\dag=\frac{1}{2}(\gamma_1+i\gamma_3),~~c_2^\dag=\frac{1}{2}(\gamma_2+i\gamma_4)\\
&c_1'^\dag=\frac{1}{2}(\gamma_1'+i\gamma_3'),~~c_2'^\dag=\frac{1}{2}(\gamma_2'+i\gamma_4')
\end{aligned}
\right.
\end{align}
With the rotation and reflection properties:
\begin{align}
\left.
\begin{aligned}
\bs{R}:&~\left(c_1^\dag,c_1'^\dag,c_2^\dag,c_2'^\dag\right)\mapsto\left(c_2^\dag,c_2'^\dag,ic_1,ic_1'\right)\\
\bs{M}:&~\left(c_1^\dag,c_1'^\dag,c_2^\dag,c_2'^\dag\right)\mapsto\left(-c_1^\dag,c_1'^\dag,-ic_2,ic_2'\right)\\
\end{aligned}
\right.
\end{align}
which can be easily verify by the rotation and reflection properties of corresponding Majorana fermions. Denote $n_j=c_j^\dag c_j$, $n_j'=c_j'^\dag c_j'$, $j=1,2$. Firstly consider the Hamiltonian with Hubbard interaction ($U>0$):
\begin{align}
H_U=U\left[(n_1-\frac{1}{2})(n_1'-\frac{1}{2})+(n_2-\frac{1}{2})(n_2'-\frac{1}{2})\right]
\end{align}
There is a 4-fold ground-state degeneracy from $(n_1,n_1')$ and $(n_2,n_2')$, which can be viewed as two spin-1/2 degrees of freedom. In order to lift this degeneracy, we define two spin-1/2 degrees of freedom as:
\begin{align}
\renewcommand\arraystretch{1.2}
\tau_j^\mu=\left(c_j^\dag,c_j'^\dag\right)\sigma^\mu\left(
\begin{array}{ccc}
c_j\\
c_j'
\end{array}
\right)
\end{align}
where $j=1,2;~\mu=x,y,z$, $\mu^j$ are Pauli matrices. We can further add the following Hamiltonian:
\begin{align}
H_J=J(\tau_1^x\tau_2^x+\tau_1^z\tau_2^z),~~J>0
\end{align}
Thus we can gap out the Majorana fermions and lift the ground state degeneracy by $H=H_U+H_J$ in a $D_4$ symmetric way.

\section{\3. Mass term for boundary Majorana fermions}
In the main text, as shown in Figs. {\color{red}3} and {\color{red}4}, we put several Majorana chains on the boundary of the system and leaves an assembly of groups of Majorana fermions on the boundary, and each group of 4 dangling Majorana fermions can be gapped out by some proper interactions and mass terms. In this section we will explain the strategy of gapping the Majorana chains on the boundary.

Consider the Majorana fermions as shown in Eq. ({\color{red}9}) in the main text. 
Firstly we consider the following Hamiltonian of interaction (where the symbol $b$ represents the Hamiltonian on the boundary):
\begin{align}
\mathcal{H}_U^b=-U\gamma_1'\gamma_1\gamma_\alpha\gamma_\delta',~U>0.
\end{align}
For ground state:
\begin{align}
\gamma_1'\gamma_1\gamma_\alpha\gamma_\delta'=1
\label{constraint B}
\end{align}
According to the Eq. ({\color{red}9}) in the main text, we can easy to verify that $\mathcal{H}_U^b$ respects the reflection symmetry. However, $\mathcal{H}_U^b$ is not enough: the eigenvalues of $\mathcal{H}_U^b$ are $\pm1$ and each eigenvalue has 2-fold degeneracy, hence there is a 2-fold ground state degeneracy. To lift this ground state degeneracy, we define the following spin-1/2 degrees of freedom:
\begin{align}
\Gamma_x=\frac{i}{2}\gamma_1'\gamma_1,~\Gamma_y=\frac{i}{2}\gamma_1'\gamma_\alpha,~\Gamma_z=\frac{i}{2}\gamma_1'\gamma_\delta'
\label{Gamma}
\end{align}
Within the constraint subspace as shown in Eq. (\ref{constraint B}), it is easy to verify that the operators defined in Eq. (\ref{Gamma}) satisfies the commutation relations of the spin-1/2 degree of freedom. Furthermore, it has the reflection symmetry properties:
\begin{align}
\bs{M}:~\left(\Gamma_x,\Gamma_y,\Gamma_z\right)\mapsto\left(-\Gamma_x,\Gamma_z,\Gamma_y\right)
\label{symmetry mass}
\end{align}
Thus we can further add a mass term to the Hamiltonian:
\begin{align}
\mathcal{H}_m^b=m(\Gamma_y+\Gamma_z),~~m>0
\end{align}
According to Eq. (\ref{symmetry mass}), $\mathcal{H}_m^b$ is symmetric under reflection. Hence the 2-fold ground state degeneracy can be lifted by the mass term $\mathcal{H}_m^b$, and this group of Majorana fermions can be gapped out by $\mathcal{H}^b=\mathcal{H}_U^b+\mathcal{H}_m^b$ in a symmetric way.

\section{\4. 2D FSPT phases with odd-fold dihedral group}
In this section, we classify 2D FSPT phases with odd-order dihedral group by explicit block-state construction, similar with the main text. We demonstrate this construction by an example: 2D FSPT phases with $D_3$ (with rotation and reflection generators $\bs{R}$ and $\bs{M}$, respectively) symmetry for the system with spinless fermions.

\begin{figure}
\centering
\includegraphics[width=0.426\textwidth]{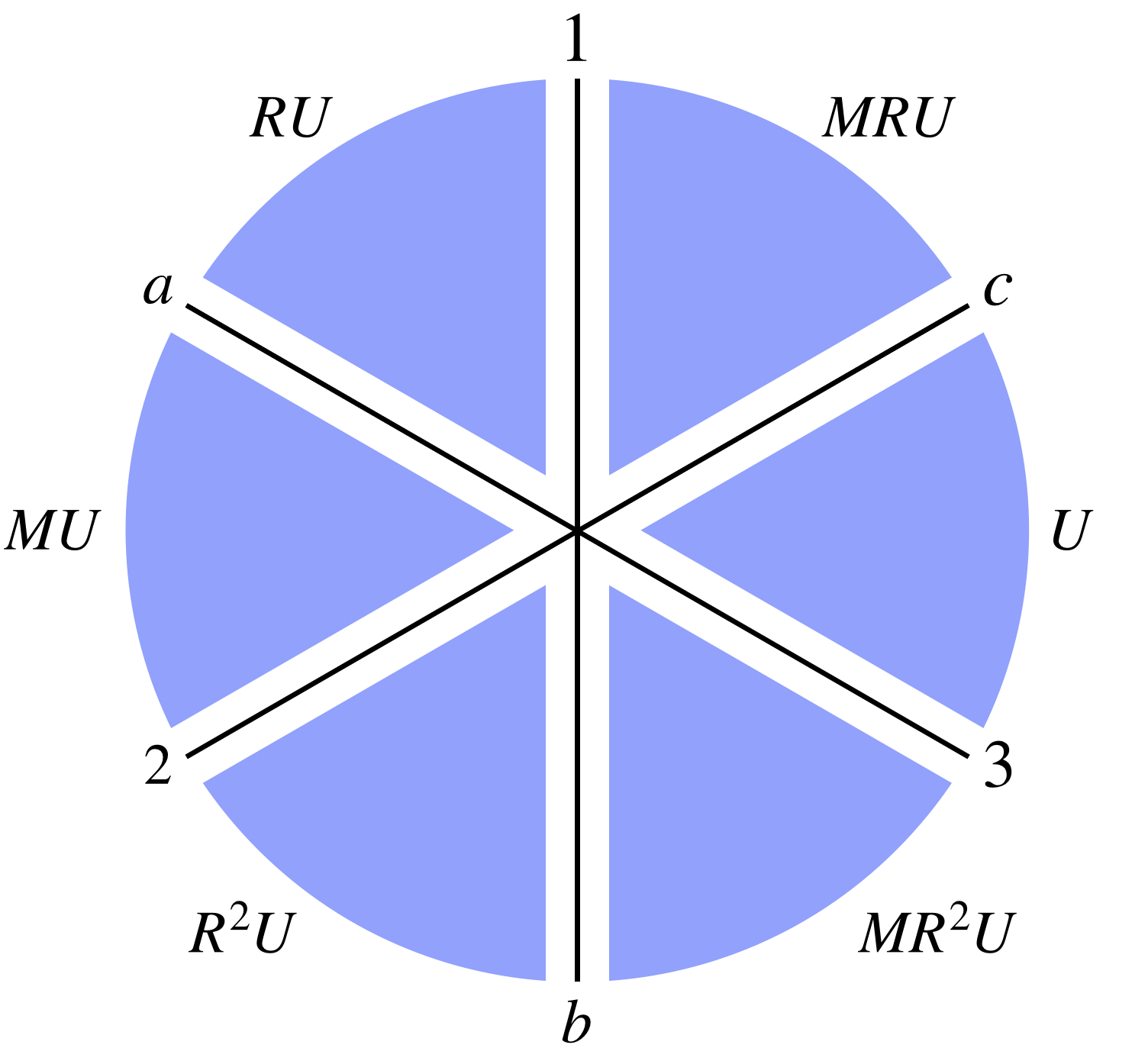}
\caption{Extended trivialization of 2D FSPT phases with $D_3$ symmetry. Here all shadowed regimes are trivialized according to Eq. (\ref{extended trivialization}).}
\label{D3 trivialization}
\end{figure}

Firstly we extended trivialize the system and encode all properties of the ground state wavefunction into some lower-dimensional block states (cf. Fig. \ref{D3 trivialization}, and we take the same notation with the main text):

\begin{align}
|\psi'\rangle=O^{\mathrm{loc}}_R|\psi\rangle=\bigotimes\limits_{g\in D_3}|T_{gU}\rangle\otimes\bigotimes\limits_{i=1,a}^{3,c}|\psi_i\rangle\otimes|\psi_{\mathrm{0D}}\rangle
\label{extended trivialization}
\end{align}

Sequentially we consider the lower-dimensional block-state decorations on 1D and 0D subsystems ($|\psi_i\rangle$ and $|\psi_{\mathrm{0D}}\rangle$, respectively).

\subsection{A. Block-state decorations}
Firstly we investigate the 1D block-state decorations. Similar with the main text, we divide these 6 semi-infinite chains into two categories: category-$\1$, $(1,2,3)$ and category-$\2$, $(a,b,c)$ because they are independent with each other under arbitrary symmetry actions of $D_3$ symmetry group (cf. Fig. \ref{D3 trivialization}). Again, for each semi-infinite chain, there are two possible root states: Majorana chain and a 1D FSPT state which can be realized by 2 copies of Majorana chains. 

For spinless fermions, there is only one possible block-state construction which can contribute a nontrivial SPT state: decorate a Majorana chain on each semi-infinite chains in both categories. It leaves 6 dangling Majorana fermions located at the 0D block, labeled by $\gamma_j,~j=1,2,...,6$, with the following symmetric properties:
\begin{align}
\left.
\begin{aligned}
&\bs{R}:\gamma_j\mapsto\gamma_{j+2},~~j\in\mathbb{Z}(\mathrm{mod}~6)\\
&\bs{M}:\gamma_j\mapsto\gamma_{8-j},~~j\in\mathbb{Z}(\mathrm{mod}~6)
\end{aligned}
\right.
\label{symmetry D3}
\end{align}
And all subscripts take the values with modulo 6. Then introduce the following Hamiltonian:
\begin{align}
\mathcal{H}=i\big(\gamma_1\gamma_4+\gamma_2\gamma_5+\gamma_3\gamma_6\big)
\label{gap}
\end{align}
It is easy to verify that Eq. (\ref{gap}) keeps invariant under arbitrary symmetry operations, according to Eq. (\ref{symmetry D3}). So Majorana chain decorations on all semi-infinite chains are compatible with all symmetries and contributes a nontrivial SPT state.

Sequentially we consider the 0D block-state decorations. $D_3$ symmetry acting on 0D block is identical with the internal symmetry $\mathbb{Z}_3\rtimes\mathbb{Z}_2$. And the number of independent irreducible representations of $\mathbb{Z}_3\rtimes\mathbb{Z}_2$ is determined by group cohomology:
\begin{align}
\mathcal{H}^1\Big[\mathbb{Z}_3\rtimes\mathbb{Z}_2,U(1)\Big]=\mathbb{Z}_2
\label{D3 0D}
\end{align}
So only nontrivial eigenvalues of reflection symmetry operation can contribute nontrivial block-state (irrelevant with the rotation subgroup) \cite{math}, and it can be trivialized by an atomic insulating state as: 
\begin{align}
|\psi\rangle_{\mathrm{0D}}=c_0^\dag c_1^\dag|0\rangle
\label{atomic}
\end{align}
here $c_0$ and $c_1$ are two complex fermions with the following reflection symmetry properties:
\begin{align}
\bs{M}:~c_0\longleftrightarrow c_1
\end{align}
Under the reflection symmetry, the atomic insulating state [cf. Eq. (\ref{atomic})] has eigenvalue $-1$:
\begin{align}
\bs{M}:~|\psi\rangle_{\mathrm{0D}}\longmapsto-|\psi\rangle_{\mathrm{0D}}
\end{align}
Furthermore, according to Ref. \onlinecite{rotation2}, a Majorana chain surrounding the 0D block is introduced to trivialize the 0D block-state with odd fermion parity, so 0D block-state decorations have no contributions to nontrivial SPT states. So the eventual classification of 2D FSPT phases with $D_3$ symmetry for spinless fermions is $\mathbb{Z}_2$.

Similar scheme of construction can be held for spin-$1/2$ fermions and obtain a trivial classification of the corresponding 2D FSPT phases with $D_3$ symmetry.

\subsection{B. Relationship with reflection SPT phases}
According to aforementioned discussions, the classifications of 2D FSPT phases with $D_3$ symmetry group coincide with the classifications of 2D FSPT phases with reflection symmetry, for both spinless and spin-$1/2$ fermions. So we discuss about the relationship between these two cases. 

The abstract group structure of $D_3$ group is $\mathbb{Z}_3\rtimes\mathbb{Z}_2$, therefore the extensions of $\mathbb{Z}_2^f$ fermion parity of $D_3$ group are labeled by the factor system of the following short exact sequence (cf. Sec. \ref{math}):
\begin{align}
0\lra\mathbb{Z}_2^f\lra G_f\lra\mathbb{Z}_3\rtimes\mathbb{Z}_2\lra0
\label{D3 extension}
\end{align}
where $G_f$ is the total symmetry group of the system. The factor systems of Eq. (\ref{D3 extension}) are labeled by 2-cocycles in:
\begin{align}
\mathcal{H}^2\left(\mathbb{Z}_3\rtimes\mathbb{Z}_2,\mathbb{Z}_2^f\right)=\mathbb{Z}_2
\label{D3 factor system}
\end{align}
And the only nontrivial 2-cocycle is solely related to the reflection subgroup of the $D_3$ group. Hence according to Eqs. (\ref{D3 0D}) and (\ref{D3 factor system}), 2D FSPT phases with $D_3$ group are only relevant with the reflection subgroup of $D_3$ symmetry group.

\begin{figure}
\centering
\includegraphics[width=0.4\textwidth]{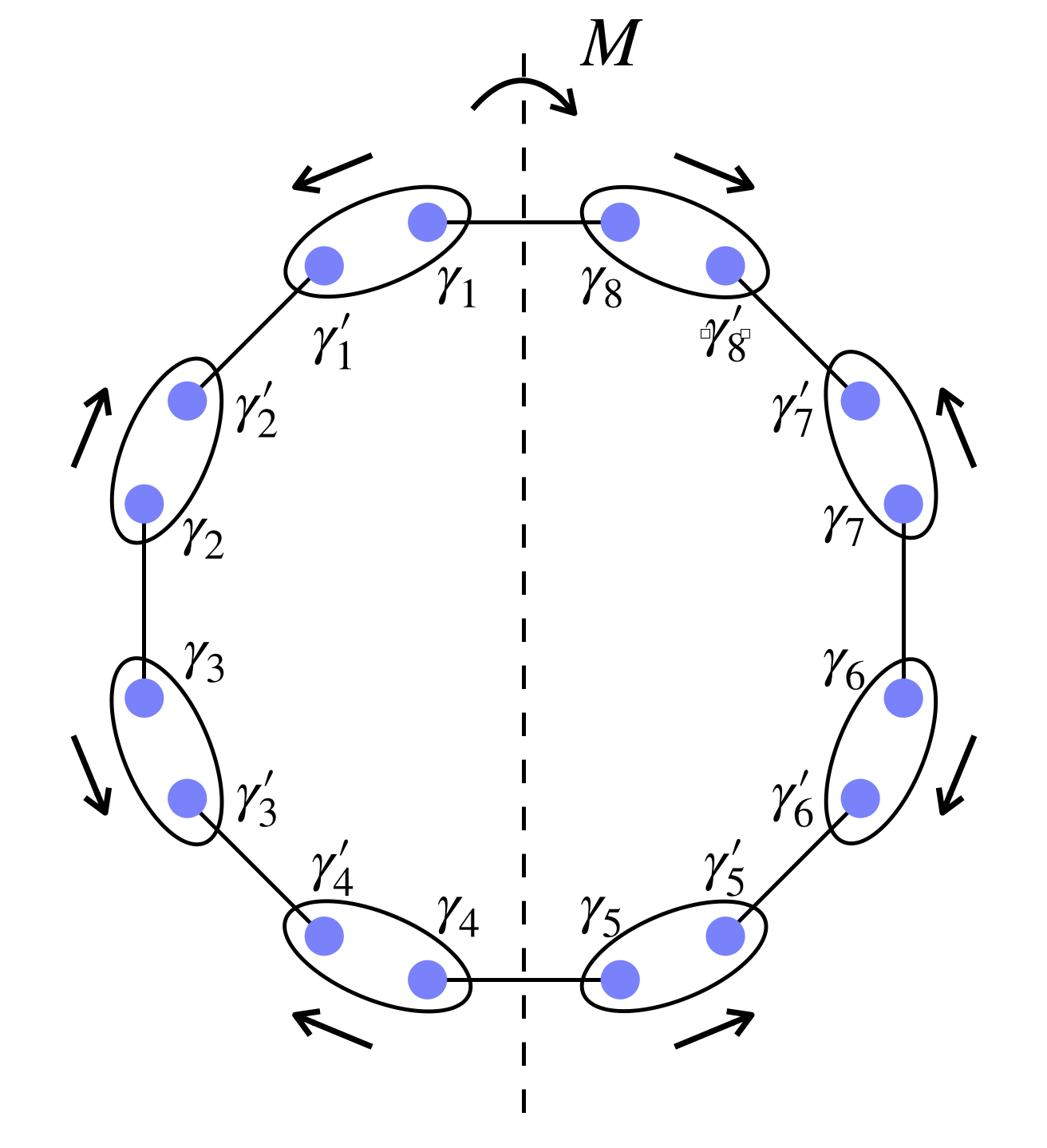}
\caption{Majorana chain surrounding the 0D block. Here each ellipse expresses an entanglement pair, each link represents a site, and the dashed line represents the reflection axis. Each arrow represents the direction of the corresponding entanglement pair.}
\label{0D}
\end{figure}

\section{\5. Trivialization of fermion parity of 0D block}
In the main text, we introduce an enclosed Majorana chain surrounding the 0D block to trivialize the nontrivial fermion parity of 0D block-state. Therefore in this section we demonstrate more details about this type of trivialization. 

The Hamiltonian of the Majorana chain surrounding the 0D block is:
\begin{align}
\mathcal{H}=i\sum\limits_{j=1}^8\gamma_j\gamma_j'
\label{0D Majorana}
\end{align}
and illustrated in Fig. \ref{0D}. It is easy to verify that Eq. (\ref{0D Majorana}) respects the full $D_4$ symmetry: the symmetry properties of these Majorana fermions are (all subscripts take the value with modulo 8):
\begin{align}
\left.
\begin{aligned}
&\bs{R}:~\gamma_j\mapsto\gamma_{j+2},~\gamma_j'\mapsto\gamma_{j+2}',~~j\in\mathbb{Z}(\mathrm{mod}~8)\\
&\bs{M}:~\gamma_j\mapsto\gamma_{9-j},~\gamma_j'\mapsto\gamma_{9-j}',~~j\in\mathbb{Z}(\mathrm{mod}~8)
\end{aligned}
\right.
\end{align}
And it is well-known that a Majorana chain ground state with periodic boundary condition (PBC) has odd fermion parity, therefore the 0D block-state with odd fermion parity can be trivialized by this Majorana chain.

\section{\6. Classification of 2D FSPT phases with internal symmetry\label{on-site dihedral}}
In this section we systematically introduce the classification of 2D FSPT phases with $\mathbb{Z}_n$ and $\mathbb{Z}_n\rtimes\mathbb{Z}_2^{\mathrm{T}}$ internal symmetry, by using the ``general group super-cohomology'' methods established in Ref. \onlinecite{2general1,2general2} and compare them with the classification of 2D FSPT phases with point group symmetries in order to conjecture the generalized ``crystalline equivalence principle''.

\subsection{A. Internal $\mathbb{Z}_n$ symmetry}
In this subsection we calculate the classification of 2D FSPT phases with internal $\mathbb{Z}_n$ symmetry, for pedagogical purpose. The total symmetry group $G_f$ is a central $\mathbb{Z}_2^f$ extension of the physical symmetry group $\mathbb{Z}_n$:
\begin{equation}
0\lra\mathbb{Z}_2^f\lra G_f\lra \mathbb{Z}_n\lra0
\label{central extension}
\end{equation}
and total number of this central extension is determined by the factor system of this short exact sequence (see Sec. \ref{math}), which is a 2-group cohomology:
\begin{align}
\mathcal{H}^2\left(\mathbb{Z}_n,\mathbb{Z}_2^f\right)=\left\{
\begin{aligned}
&\mathbb{Z}_1,~~\text{$n$ is odd}\\
&\mathbb{Z}_2,~~\text{$n$ is even}
\end{aligned}
\right.
\label{rotation factor system}
\end{align}
so we shall investigate the classifications of odd/even $n$ separately. 
\subsubsection{1. $n\in$ odd}
For $n\in$ odd case, according to Eq. (\ref{rotation factor system}), there is only one possible (trivial) extension of fermion parity, which is labeled by $\mathbb{Z}_2^f\times\mathbb{Z}_n$. The classification data are:
\begin{align}
(n_1,n_2,\nu_3)&\in\mathcal{H}^1(\mathbb{Z}_n,\mathbb{Z}_2)\times\mathcal{H}^2(\mathbb{Z}_n,\mathbb{Z}_2)\times\mathcal{H}^3\left[\mathbb{Z}_n,U(1)\right]\nonumber\\
&=\mathbb{Z}_1\times\mathbb{Z}_1\times\mathbb{Z}_n
\end{align}
and there are no obstructions or trivializations. So the ultimate classification of 2D FSPT phases with internal $\mathbb{Z}_n$ symmetry ($n\in$ odd) is $\mathbb{Z}_n$.

\subsubsection{2. $n\in$ even}
For $n\in$ even case, according to Eq. (\ref{rotation factor system}), there are two possible extensions of fermion parity, which are labeled by $\mathbb{Z}_2^f\times\mathbb{Z}_n$ and $\mathbb{Z}_{2n}^f$ ($g^n=\pm1,~g\in\mathbb{Z}_n$). The classification data are:
\begin{align}
(n_1,n_2,\nu_3)&\in\mathcal{H}^1(\mathbb{Z}_n,\mathbb{Z}_2)\times\mathcal{H}^2(\mathbb{Z}_n,\mathbb{Z}_2)\times\mathcal{H}^3\left[\mathbb{Z}_n,U(1)\right]\nonumber\\
&=\mathbb{Z}_2\times\mathbb{Z}_2\times\mathbb{Z}_n
\end{align}
For $\mathbb{Z}_n\times\mathbb{Z}_2^f$ case, there are no obstructions and trivializations. So the ultimate classification of 2D FSPT phases with internal $\mathbb{Z}_n$ symmetry ($n\in$ even) is $\mathbb{Z}_2^2\times\mathbb{Z}_n$. For $\mathbb{Z}_{2n}^f$ case, all nontrivial $n_1$ and $n_2$ are obstructed, and half of nontrivial $\nu_3$ are trivialized by anomalous symmetry protected topological (ASPT) phases \cite{anomalousSPT} characterized by $(-1)^{\omega_2\smile n_1}$, where $\omega_2$ characterizes the factor system of the central extension Eq. (\ref{central extension}). All results are summarized in the main text, for different $\mathbb{Z}_n$ and systems with spinless/spin-$1/2$ fermions (different extensions of fermion parity).

\subsection{B. Internal $\mathbb{Z}_n\rtimes\mathbb{Z}_2^{\mathrm{T}}$ symmetry}
In this subsection we calculate the classification of 2D FSPT phases with internal $\mathbb{Z}_n\rtimes\mathbb{Z}_2^{\mathrm{T}}$ symmetry, where $\mathbb{Z}_2^{\mathrm{T}}$ is antiunitary time-reversal symmetry. The total symmetry group $G_f$ is a central extension of fermion parity of physical symmetry:
\begin{align}
0\lra\mathbb{Z}_2^f\lra G_f\lra \mathbb{Z}_n\rtimes\mathbb{Z}_2^{\mathrm{T}}\lra0
\label{dihedral extension}
\end{align}
and total number of this central extension is determined by the factor system of this short exact sequence:
\begin{align}
\mathcal{H}^2\left(\mathbb{Z}_n\rtimes\mathbb{Z}_2^{\mathrm{T}},\mathbb{Z}_2^f\right)=\left\{
\begin{aligned}
&\mathbb{Z}_2,~~\text{$n$ is odd}\\
&\mathbb{Z}_2^3,~~\text{$n$ is even}
\end{aligned}
\right.
\label{dihedral factor system}
\end{align}
So we shall investigate the classifications of odd/even $n$ separately.

\subsubsection{1. $n\in$ odd}
For $n\in$ odd case, according to Eq. (\ref{dihedral factor system}), there are 2 possible extensions of fermion parity (actually the fermion parity only extends the time-reversal subgroup of $G_b=\mathbb{Z}_n\rtimes\mathbb{Z}_2^{\mathrm{T}}$ group for $n\in$ odd case), which are labeled by $\mathcal{T}^2=\pm1$ (which corresponds to spinless and spin-$1/2$ fermions). The classification data are:
\begin{align}
(n_1,n_2,\nu_3)&\in\mathcal{H}^1(G_b,\mathbb{Z}_2)\times\mathcal{H}^2(G_b,\mathbb{Z}_2)\times\mathcal{H}^3\left[G_b,U_{\mathrm{T}}(1)\right]\nonumber\\
&=\mathbb{Z}_2\times\mathbb{Z}_2\times\mathbb{Z}_1
\end{align}
for $\mathcal{T}^2=1$ case (spinless fermions), all nontrivial $n_1$ and $n_2$ are obstructed, so the ultimate classification of this case is trivial. For $\mathcal{T}^2=-1$ case (spin-$1/2$ fermions), nontrivial $n_2$ (which corresponds to the complex fermion decoration) is trivialized, therefore the ultimate classification of this case is $\mathbb{Z}_2$. All results are summarized in the main text.

\subsubsection{2. $n\in$ even}
For $n\in$ even integer case, according to Eq. (\ref{dihedral factor system}), there are 8 possible extensions of fermion parity of the $\mathbb{Z}_n\rtimes\mathbb{Z}_2^{\mathrm{T}}$ group, which are labeled by 3 indices $(\mathcal{T}^2,g^{n},(\mathcal{T}g^{n/2})^2)=(\pm1,\pm1,\pm1)$ as abbreviated 2-cocycles, where $\mathcal{T}\in\mathbb{Z}_2^{\mathrm{T}}$ and $g\in\mathbb{Z}_n$ are generators of the corresponding symmetry group. We demonstrate only 2 of them for simplicity: $\mathcal{T}^2=g^{n}=(\mathcal{T}g^{n/2})^2=\pm1$ (which correspond to systems with spinless and spin-$1/2$ fermions), and similar arguments can be held to all other cases with different extensions of fermion parity. The classification data are:
\begin{align}
(n_1,n_2,\nu_3)&\in\mathcal{H}^1(G_b,\mathbb{Z}_2)\times\mathcal{H}^2(G_b,\mathbb{Z}_2)\times\mathcal{H}^3\left[G_b,U_{\mathrm{T}}(1)\right]\nonumber\\
&=\mathbb{Z}_2^2\times\mathbb{Z}_2^3\times\mathbb{Z}_2^2
\end{align}
And the explicit expressions of cocycles for $\mathcal{H}^1(G_b,\mathbb{Z}_2)=\mathbb{Z}_2^2$ and $\mathcal{H}^2(G_b,\mathbb{Z}_2)=\mathbb{Z}_2^3$ are also be presented as follows. Let $a=x^{a_x}y^{a_y}~(0\leq a_x<n,0\leq a_y\leq1)$ be an element of $D_n$. Then $\mathcal{H}^1(D_n,\mathbb{Z}_2)=\mathbb{Z}_2^2$ is generated by:
\begin{align}
\left.
\begin{aligned}
&n_1(a)=a_x~(\mathrm{mod}~2)\\
&n_1(a)=a_y~(\mathrm{mod}~2)
\end{aligned}
\right.
\end{align}
and $\mathcal{H}^2(D_n,\mathbb{Z}_2)=\mathbb{Z}_2^3$ is generated by:
\begin{align}
\left.
\begin{aligned}
&n_2(a,b)=a_xb_x~(\mathrm{mod}~2)\\
&n_2(a,b)=a_yb_y~(\mathrm{mod}~2)\\
&n_2(a,b)=\left\lfloor\frac{[a_x+(-1)^{a_y}b_x]_{2n}}{n}\right\rfloor
\end{aligned}
\right.
\label{2-cocycles}
\end{align}
where we define $[x]_n=x~(\mathrm{mod}~n)$ and $\lfloor x\rfloor$ as the greatest integer less than or equal to $x$.

For $\mathcal{T}^2=g^{n}=(\mathcal{T}g^{n/2})^2=-1$ case (system with spin-$1/2$ fermions), only one nontrivial $n_2$ is obstruction free (the last formula in Eq. (\ref{2-cocycles}), which correspond to the complex fermion decoration of 2D FSPT phases), all other nontrivial $n_1$ and $n_2$ are obstructed. Furthermore, all nontrivial 3-cocycle $\nu_3$ are trivialized by nontrivial $(-1)^{\omega_2\smile n_1}$ (ASPT phases on the boundary) which is not a 3-coboundary. So the ultimate classification of this case is $\mathbb{Z}_2$. 

For $\mathcal{T}^2=g^{n}=(\mathcal{T}g^{n/2})^2=1$ case, all nontrivial $n_1$ and $n_2$ are obstructed, but there is no trivialization on $\nu_3$ because the factor system $\omega_2\in\mathcal{H}^2(G_b,\mathbb{Z}_2^f)$ is trivial. So the ultimate classification of this case are contributed solely from $\nu_3\in\mathcal{H}^3[G_b,U_{\mathrm{T}}(1)]=\mathbb{Z}_2^2$. All results for these cases are summarized in the main text.

\providecommand{\noopsort}[1]{}\providecommand{\singleletter}[1]{#1}%
%


\begin{thebibliography}{56}
\bibitem [{pri()}]{private}%
  \BibitemOpen
  \href@noop {} {\bibinfo  {journal} {The space group of 3D iron selenide is
  $P4/nmm$, and according to recent STM experiments, there is no nematicity for
  monolayer iron selenide. Introduce some defect to break translational
  symmetry to realize $D_4$ symmetry}\ }\BibitemShut {NoStop}%
\end{thebibliography}

\end{document}